\documentclass[aps,prl,twocolumn,groupedaddress, showpacs]{revtex4}

\usepackage{graphicx}

\begin{document}


\title{Motion-Induced Magnetic Resonance of Rb Atoms \\
in a Periodic Magnetostatic Field}


\author{A. Hatakeyama}
 \email[Electronic address: ]{hatakeya@phys.c.u-tokyo.ac.jp}
\author{Y. Enomoto}
\author{K. Komaki}
\author{Y. Yamazaki}
\altaffiliation[Also at ]{Atomic Physics Laboratory, RIKEN, Wako, Saitama 351-0198, Japan.}
\affiliation{
Graduate School of Arts and Sciences,
University of Tokyo, Komaba, Tokyo 153-8902, Japan
}



\date{\today}

\begin{abstract}
We demonstrate that transitions between Zeeman-split sublevels of Rb atoms are resonantly induced by the motion of the atoms (velocity: $\sim100$~m/s) in a periodic magnetostatic field (period: 1~mm) when the Zeeman splitting corresponds to the frequency of the magnetic field experienced by the moving atoms. A circularly polarized laser beam polarizes Rb atoms with a velocity selected using the Doppler effect and detects their magnetic resonance in a thin cell, to which the periodic field is applied with the arrays of parallel current-carrying wires.
\end{abstract}


\maketitle

By applying an electromagnetic (EM) wave to an atom at rest, 
one can resonantly induce
transition between two atomic energy levels whose
frequency difference coincides with the EM wave frequency.
This resonance transition is one of the most extensively studied
and widely utilized phenomena
in atomic physics and also in other various fields.
Also in the reversed configuration, that is, by an atom moving in a static periodic field,
an internal transition can be made 
when the periodic perturbation experienced
by the atom has a frequency equal to the transition frequency.
The resonance transition of this kind, quite simple in principle and applicable to any particle with internal states, however, has been clearly demonstrated only
in the phenomenon called ``resonant coherent excitation'' (RCE)~\cite{Dat78}
using channeled fast ion beams in periodic
electric fields of crystals.
RCE has attracted
much attention~\cite{Kra96,Gar04}
since the first
proposal by Okorokov~\cite{Oko65}.
Recent progress made by using the high-energy
beams of highly charged ions shows the possibility of
high resolution spectroscopy of
highly charged ions in an x-ray region of keV or $10^{18}$~Hz~\cite{Azu99}.
The order of this transition frequency is determined by an ion velocity  ($\sim 10^8$~m/s) and a lattice constant ($\sim 10^{-10}$~m).

We report in this Letter on the resonance transition that is based on
the same
principles as RCE but is induced by a different type of interaction in a quite
different energy range of neV or $10^5$~Hz:
magnetic resonance between the Zeeman sublevels of 
Rb atoms with a $z$ 
velocity component $v_z$ of $\sim 100$~m/s in a
static magnetic field that is periodic in the $z$ direction
(period: $a=1$~mm).
This resonance transition, called motion-induced resonance in this Letter,
was observed in a thin cell containing Rb vapor,
with the periodic magnetic
field applied with the arrays of parallel current-carrying
wires sandwiching the cell. 
The atoms with a velocity selected using the Doppler effect
were
polarized by optical pumping with circularly polarized laser light slightly detuned
from the $D_2$ line.
The magnetic resonance between the
ground state sublevels was optically
detected with the same laser beam.
We confirmed that the resonance occurs when the Zeeman splitting frequency
coincides with $v_z/a$, namely the frequency of the field experienced by the
moving atoms. We successfully obtained
the resonance spectra very similar to those of
standard rf magnetic resonance.
Our clear demonstration of motion-induced resonance shows that
experiments 
using slow atoms and artificial periodic fields
are useful to study the fundamental dynamics of this
kind of resonance compared to RCE experiments, which involve complicated
atomic processes in solid.
We also consider that this experiment is the first step
to extend the study to well-controlled
atoms in the fields of precise- and small-period structures with the help of
advancing surface nanofabrication technologies
and progressing atom control and
manipulation techniques, particularly near surfaces,
such as atom chips~\cite{Fol00} and quantum reflection~\cite{Shi01}. 
These studies will reveal a new aspect of motion-induced resonance,
and may lead to the development of
unique techniques alternative to EM wave methods to control
the internal states and also the associated motional states
 (see the next paragraph) of atoms (and any particle with internal states)
near surfaces.

We first discuss
the basic properties of motion-induced resonance from a 
perspective of the momentum and energy of an atom
in analogy to atom-EM-wave resonance
to get a better understanding
rather than the simple picture that the resonance transition occurs when
the frequency of the field experienced by an atom coincides with the
transition frequency.
When the internal state of an atom is excited in an EM standing wave,
the momentums and energies of the atom before and after excitation are
related as
\begin{equation}
mv'_z=mv_z \pm h/\lambda, \quad
E'+mv'^{2}/2=E+mv^2/2+h\nu,
\label{eq:photon}
\end{equation}
where $v_z(v'_z) (>0)$, $E(E')$, and $v(v')$ are, respectively, the velocity in the $z$ direction, the internal
energy of the atom, and the speed of the atom with a mass of $m$ before (after) 
excitation
by the EM wave 
with a frequency of $\nu$ and a wavelength of $\lambda$
(note $v^2=v^2_x+v^2_y+v^2_z$).
$h$ is Planck's constant.
The sign $+$ ($-$) in the momentum relation corresponds to the excitation
induced by one of the counterpropagating waves that is running
in the $+z$ ($-z$) direction.
Neglecting the small term $h^2/(2m{\lambda}^2)$,
the so-called recoil energy~\cite{Mey91},
one can deduce from Eq.~(\ref{eq:photon}) the well-known
resonance condition
$E'-E=\Delta E=h\nu \mp hv_z/\lambda$, where
$hv_z/\lambda$ represents the Doppler shift.
Note that although
$h/\lambda$ and $h\nu$ are regarded
as the momentum and the energy of a photon, respectively, 
the momentum and energy relations in
Eq.~(\ref{eq:photon}) can be derived from the classical
description of field periodicity and oscillation, repsectively~\cite{Mey91}. 
Therefore in the case of motion-induced resonance
it is justified that, in an inertial 
frame where the static periodic field is at rest, one sets $\nu=0$
while keeping $\lambda=a$ in Eq.~(\ref{eq:photon}).
The corresponding relations are
\begin{equation}
mv'_z=mv_z \pm h/a, \quad
E^{'}+mv'^2/2=E+mv^2/2.
\label{eq:MIR}
\end{equation}
One has to take the lower sign in the case $\Delta E > 0$.
The resonance condition $\Delta E=hv_z/a$
is then deduced.
From this discussion it can be said that the transition is caused purely
by the Doppler effect, and the
internal energy is increased at the expense of the kinetic energy.
In addition, as a consequence of the field periodicity,
the atom momentum should change by
$h/a$ in the transition (note that this momentum change is transfered
to the periodic structure that produces the periodic field). 
This fact, although
not having been verified experimentally,
may be used
to control the motional state of atoms by motion-induced
resonance.

\begin{figure}
\includegraphics[width=8cm]{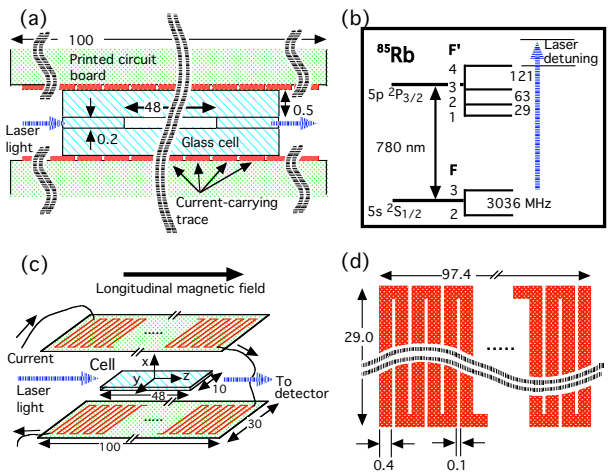}
\caption{\label{fig1} (color online). (a) Schematic of the cross section of the thin cell and sandwiching printed circuit boards (PCBs). The dimensions are in millimeter. (b) Relevant energy levels of $^{85}$Rb. The laser detuning is also shown. (c) Current direction on the PCBs. The origin of the coordinate system, which is used in Fig.~\ref{fig2}, is at the center of the cell volume. (d) Current-carrying trace on a PCB. The copper trace is about 35~$\mu m$ thick, covered with a thiner insulating film.}
\end{figure}

The experiment was performed with a thin quartz cell (inner dimension: 
48 $\times$ 10 $\times$ 0.2~mm), whose cross section is depicted
in Fig.~\ref{fig1}(a). The evacuated cell contains Rb vapor with
a density of $\sim10^9$~cm$^{-3}$
at room temperature, where the Doppler broadening
for the Rb $D_2$ transition at 780~nm is about 
250~MHz (half width at half maximum: HWHM).
The natural linewidth
of the $D_2$ line is 3~MHz (HWHM).
An external cavity laser diode, whose linewidth
was estimated to be 1~MHz,
was frequency stabilized to the F=3
$\rightarrow$ F$'$=4 transition of the $D_2$ line of $^{85}$Rb 
(see Fig.~\ref{fig1}(b)).
The uncertainty of the laser frequency was at most a few MHz.
The laser frequency was then shifted (see Fig.~\ref{fig1}(b))
with an acousto-optic frequency shifter 
in the range of 123-203 MHz
to select an atom velocity in the range of
$v_z=96$ to 
$158$~m/s through the F=3 $\rightarrow$ F$'$=4 transition. 
The laser beam was finally made right-circularly polarized and
irradiated the whole volume of the
narrow cell gap to polarize the F=3 ground state of
atoms 
with a velocity component $v_z$ in the laser direction. 
Note that atoms with 
a velocity $v_z+94$ or $v_z+144$~m/s were also
polarized through the transition F=3 $\rightarrow$ F$'$=3 or F$'$=2.
However, they are fewer than atoms with a velocity $v_z$, and 
also tend to become off resonant with the laser due to
hyperfine pumping to the F=2 ground state via the excited states.
Therefore they should not  significantly contribute to the signal.
A periodic magnetostatic field (period: $a=1$~mm) was applied to the cell with two printed circuit boards (PCBs) 
(100 $\times$ 30~mm) sandwiching the cell. The current direction and the current-carrying trace
on a PCB are shown in Figs.~\ref{fig1}(c), (d), respectively.
The details of the produced periodic field
is described in the next paragraph.
A longitudinal magnetic field applied along the laser beam with a set of Helmholtz coils (40~cm diam)
was slowly scanned to cause the Zeeman splitting 
by 4.67~kHz per $\mu$T between the adjacent sublevels of the F=3 state.
The resonant transitions between the sublevels occurred
when the splitting frequency coincided with the frequency
of the field experienced by the atoms under observation.
The resonance was detected by monitoring the transmitted
laser intensity, which increased when the polarization was 
destroyed at resonance. This increase corresponded to
typically $10^{-5}-10^{-6}$
variation of the laser beam intensity. 
For better detection sensitivity, we employed a lock-in detection scheme by switching on and off the periodic field at 4 kHz, and further averaged the lock-in signals.
The cell, PCBs and coils were enclosed in a magnetic shield (65 $\times$ 65 $\times$ 65~cm), which reduced the Earth's and other environmental 
magnetic fields to the order of 0.1~$\mu$T.
To perform a standard
rf magnetic resonance experiment for comparison with motion-induced resonance, the PCBs were removed and
then resonance profiles were recorded
as a transverse oscillating magnetic field was applied with
another set of Helmholtz coils (15~cm diam).

\begin{figure}
\includegraphics[width=8cm]{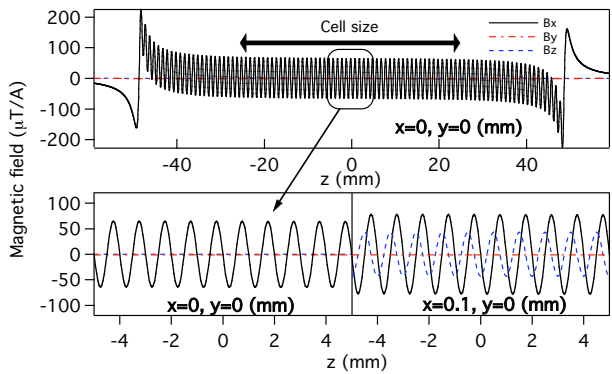}
\caption{\label{fig2} (color online). Calculated magnetic field produced by unit PCB current. 
The coordinate system is defined in Fig.~\ref{fig1}(c).
The cell is located in the center of the PCBs and the current flows on the surfaces at
$x=\pm 0.65$~mm. The upper graph shows the whole picture
of the produced field at $x=y=0$~mm (cell center) as a function of $z$.
The lower left shows the details around $z=0$~mm, while the lower right shows the details at $x=0.1$~mm, $y=0$~mm (cell surface).}
\end{figure}

The basic behavior of the field produced by the PCBs in the cell is
described analytically, using the coordinate system
defined in Fig.~\ref{fig1}(c), as
\begin{eqnarray}
&B_x = B_0 \exp (-kd) \cosh (kx) \sin (kz), \ \ \
B_y=0, \nonumber \\ 
&B_z = B_0 \exp(-kd) \sinh (kx) \cos (kz).
\label{eq:field2}
\end{eqnarray}
$B_0$ is a constant proportional to the current, and $k=2\pi/a$.
The current flows on the surfaces located at $x=\pm d$;
$d$ was measured to be 0.65~mm.
Figure~\ref{fig2} shows the numerically calculated periodic magnetic field
produced by the actual PCB current. 
At the center of the cell ($x=y=0$~mm), the $x$ component of the
field $B_x$ behaves as $\sin(kz)$ with an amplitude
of 65~$\mu$T/A, although its baseline declines
slightly along the $z$ direction due to the finite number of current wires~\cite{correction_wires} as seen in the upper graph in
Fig.~\ref{fig2}. $B_z$ is equal to zero, while $B_y$ has a small value of $1\sim2$~$\mu$T/A, which is produced by the current flowing in the $z$ direction on the PCB edges at $y \simeq \pm15$~mm. 
Near the cell surfaces at $x=\pm 0.1$~mm,
$B_x$ is larger by 20\% than at the center as seen in the
lower-right graph in Fig.~\ref{fig2},
while $B_z$ oscillates as well with an amplitude of 43~$\mu$T/A.
This longitudinal 
$B_z$ field experienced by moving atoms is so rapidly oscillating, 
however, that it does not affect resonance profiles in the experiment.
Note that the sandwiching configuration of the PCBs reduces a
strong dependence of the field strength on $x$ around $x=0$, contrary to an exponentially ($1/e$
constant: $a/(2\pi)$) decreasing field with increasing distance from a
single periodic source~\cite{Hin99}.
Finally, near the cell edges at $y=\pm 5$~mm the produced field is
basically the same as described above. 
As a summary, the produced periodic field can essentially approximate
to a sinusoidal field oscillating in the $x$ direction
with a constant amplitude over the cell.
It is therefore expected to obtain motion-induced
resonance profiles similar
to rf resonance ones.
We finally note that the calculated field strength has an
uncertainty of about $\pm 10$\%, which is mainly due to an uncertainty
in the position of the PCB current ($d = 0.65\pm0.02$~mm) .

\begin{figure}
\includegraphics[width=6cm]{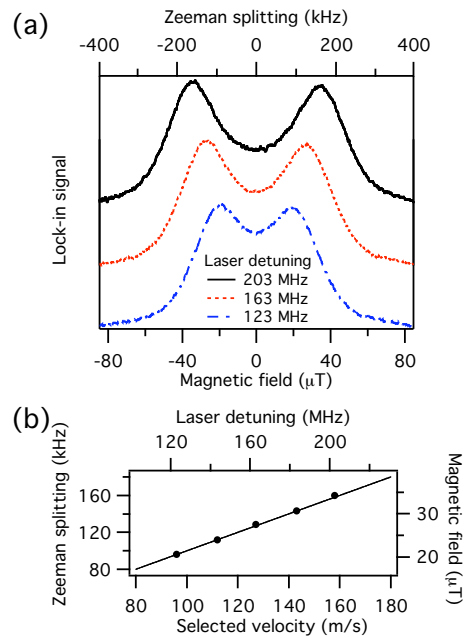}
\caption{\label{fig3} (color online). (a) Motion-induced resonance spectra obtained for various laser detunings.
The laser intensity is 140~$\mu$W/cm$^2$ with $\pm$10\% uncertainty. The PCB current is 0.20~A
and the estimated periodic field
amplitude at the cell center is 13~$\mu$T.
The spectra have arbitrary offsets for better display. The signal scales are
roughly the same.
(b) Zeeman splitting $f$ at which the resonance peak is centered as a function of selected velocity $v_z$. The right and top scales show the corresponding longitudinal magnetic field and laser detuning. The line represents the relation $f=v_z/a$ ($a=1$~mm).}
\end{figure}
\begin{figure}
\includegraphics[width=6.5cm]{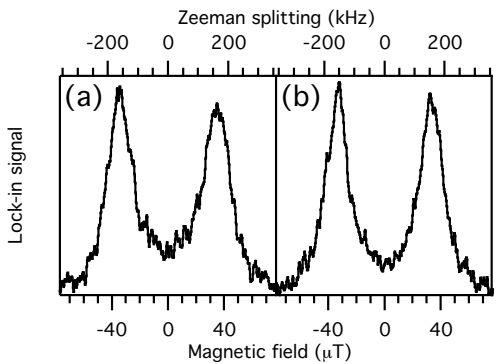}%
\caption{\label{fig4} (a) Motion-induced resonance spectrum
obtained for a laser detuning of 203~MHz
and hence a selected velocity of 158~m/s.
The PCB current is 0.08~A and the estimated periodic magnetic field
amplitude at the cell center is 5.2~$\mu$T. (b) 
rf resonance spectrum obtained
for an rf frequency of 156~kHz and amplitude of 5.0~$\mu$T.
The laser detuning is the same as that in (a).
In (a) and (b) the laser intensities are almost equal,
$40~\mu$W/cm$^2$ with $\pm 20$\% uncertainties. 
The signal scales
are roughly the same.}
\end{figure}

Figure \ref{fig3}(a) shows motion-induced magnetic resonance spectra recorded as a function of the longitudinal magnetic field from about $-80$~$\mu$T to $+80$~$\mu$T. The laser frequency detunings are 203~MHz, 163~MHz, and
123~MHz,
corresponding to selected velocities of 158~m/s, 127~m/s, and 96~m/s, respectively.
The two resonance peaks are clearly observed,
almost symmetric with respect to the
zero magnetic field.
One may notice the slightly larger 
resonance signals in the negative longitudinal field than in the
positive field, noticeable also in rf resonance spectra 
(see Fig.~\ref{fig4}(b)). We consider that this is attributed to 
a small difference in the laser absorption probability due to the Zeeman shifts of sublevels.
As the laser detuning (and hence the selected velocity) decreases, the
resonance peaks move toward the zero magnetic field (and hence the zero
Zeeman splitting) as seen in Fig.~\ref{fig3}(a).
We derived with an uncertainty of a few kHz
the sublevel splitting $f$ at which the resonance peak was centered
with the help of Lorentzian fittings, and
plot it  as a function of selected velocity $v_z$
in Fig.~\ref{fig3}(b). As clearly seen,
the resonance frequency is
proportional to the atom velocity as $f=v_z/a$ $(a=1~$mm), 
which is the resonance condition of motion-induced
resonance.

The spectral width was found to be
mainly determined by the light
intensity and the periodic field strength
from measurements performed by changing each parameter
independently.
Figure~\ref{fig4}(a) shows
narrower linewidth for a weaker laser intensity
and PCB current than in 
Fig.~\ref{fig3}(a).
The linewidth was still decreasing with decreasing laser
intensity and PCB current in the ranges of the intensity and current
of Fig.~\ref{fig4}(a), although we could not reduce them
further due to the limit of detection sensitivity in the present
experimental setup.
The width (HWHM) of the spectrum in Fig.~\ref{fig4}(a) is
56~kHz (12~$\mu$T), 
corresponding to an effective transverse spin relaxation time of
3~$\mu$s. This relaxation time is of the same order as
the mean free time between collisions of atoms with the cell surfaces, where the spin polarization is supposed to be completely
destroyed.
It is interesting to point out that the spectral width should become much narrower with decreasing laser intensity than that determined by the mean free time, because of selective polarization and detection of atoms that have long free flight times (and hence have long spin relaxation times)~\cite{Bri99}.

We show in Fig.~\ref{fig4}(b) a magnetic resonance spectrum recorded in the standard magnetic resonance experiment, namely recorded with a transverse rf magnetic field oscillating at 156 kHz, for
the laser intensity and field amplitude similar to the case of Fig.~\ref{fig4}(a).
The laser detuning is the same as that in Fig.~\ref{fig4}(a), although
the selected velocity dependence is negligible in rf resonance.
As seen in this figure, the shape of the spectrum
is basically identical to that of the motion-induced resonance as we have
expected, while its width is somewhat smaller (50~kHz HWHM). 
This small difference may probably be attributed to
additional factors
that increase the resonance
width in motion-induced resonance in our experimental setup. One factor
is the finite width in selected velocity, originating from the natural linewidth and the laser-power broadening of the
$D_2$ absorption line. The non-periodic transverse
magnetic field produced by the PCB current is another factor, about a few $\mu$T/A in the $x$ and $y$ directions. The current
in the wires connecting the two PCBs and the PCBs to a power supply
also produces a transverse field of the same order.

In conclusion, we have clearly demonstrated that magnetic transitions are
resonantly induced in an artificial periodic
magnetic structure by the motion of Rb atoms having thermal
energy at room temperature.
We have confirmed that the magnetic resonance occurs 
when the Zeeman splitting of the ground state of Rb atoms
corresponds to the atom velocity 
divided by the field period.
Resonance profiles quite similar 
to those of standard rf magnetic resonance have been obtained. 
We finally emphasize that motion-induced resonance, 
whose principles have been demonstrated
only in so-called resonant coherent excitation by using channeled fast 
ion beams in crystal fields, is a general
phenomenon that occurs for various scales of field periods and kinetic/transition energies of particles, and also for various types of particles and interactions.
Our demonstration opens the door to a new class of experiments related to
this resonance for slow atoms in artificial periodic fields.

This work was supported by the Grant-in-Aid for Scientific Research from the Ministry of Education, Culture, Sports, Science and Technology, Japan, and by Matsuo Foundation. We also thank
a grant-in-aid from The 21st Century COE (Center of Excellence)
program (Research Center for Integrated Science) of the same ministry.


\end{document}